\documentclass[aps,prl,twocolumn,showpacs,superscriptaddress]{revtex4}
\usepackage{graphicx}
\usepackage{color}

\begin{document}
\title{Single-layer MoS$_2$ on Au(111): band gap renormalization and substrate interaction}
\author{Albert Bruix}
\affiliation{Department of Physics and Astronomy, Interdisciplinary Nanoscience Center (iNANO), Aarhus University, 8000 Aarhus C, Denmark}
\author{Jill A. Miwa}
\affiliation{Department of Physics and Astronomy, Interdisciplinary Nanoscience Center (iNANO), Aarhus University, 8000 Aarhus C, Denmark}
\author{Nadine Hauptmann}
\affiliation{Institute for Molecules and Materials, Radboud University, 6500 GL Nijmegen, The Netherlands}
\author{Daniel Wegner}
\affiliation{Institute for Molecules and Materials, Radboud University,  6500 GL Nijmegen, The Netherlands}
\author{S\o ren Ulstrup}
\affiliation{Department of Physics and Astronomy, Interdisciplinary Nanoscience Center (iNANO), Aarhus University, 8000 Aarhus C, Denmark}
\author{Signe S. Gr\o nborg}
\affiliation{Department of Physics and Astronomy, Interdisciplinary Nanoscience Center (iNANO), Aarhus University, 8000 Aarhus C, Denmark}
\author{Charlotte E. Sanders}
\affiliation{Department of Physics and Astronomy, Interdisciplinary Nanoscience Center (iNANO), Aarhus University, 8000 Aarhus C, Denmark}
\author{Maciej Dendzik}
\affiliation{Department of Physics and Astronomy, Interdisciplinary Nanoscience Center (iNANO), Aarhus University, 8000 Aarhus C, Denmark}
\author{Antonija Grubi\v{s}i\'c \v{C}abo}
\affiliation{Department of Physics and Astronomy, Interdisciplinary Nanoscience Center (iNANO), Aarhus University, 8000 Aarhus C, Denmark}
\author{Marco Bianchi}
\affiliation{Department of Physics and Astronomy, Interdisciplinary Nanoscience Center (iNANO), Aarhus University, 8000 Aarhus C, Denmark}
\author{Jeppe V. Lauritsen}
\affiliation{Department of Physics and Astronomy, Interdisciplinary Nanoscience Center (iNANO), Aarhus University, 8000 Aarhus C, Denmark}
\author{Alexander A. Khajetoorians}
\affiliation{Institute for Molecules and Materials, Radboud University,  6500 GL Nijmegen, The Netherlands}
\author{Bj\o rk Hammer}
\affiliation{Department of Physics and Astronomy, Interdisciplinary Nanoscience Center (iNANO), Aarhus University, 8000 Aarhus C, Denmark}
\author{Philip Hofmann}
\email{philip@phys.au.dk}
\affiliation{Department of Physics and Astronomy, Interdisciplinary Nanoscience Center (iNANO), Aarhus University, 8000 Aarhus C, Denmark}

\date{\today}
\begin{abstract}
The electronic structure of epitaxial single-layer MoS$_2$ on Au(111) is investigated by angle-resolved photoemission spectroscopy, scanning tunnelling spectroscopy, and first principles calculations. While the band dispersion of the supported single-layer is close to a free-standing layer in the vicinity of the valence band maximum at  $\bar{K}$ and the calculated electronic band gap on Au(111) is similar to that calculated for the free-standing layer, significant modifications to the band structure are observed at other points of the two-dimensional Brillouin zone: At $\bar{\Gamma}$, the valence band maximum has a significantly higher binding energy than in the free MoS$_2$ layer and the expected spin-degeneracy of the uppermost valence band at the $\bar{M}$ point cannot be observed. These band structure changes are reproduced by the calculations and can be explained by the detailed interaction of the out-of-plane MoS$_2$ orbitals with the substrate.
\end{abstract}
\pacs{73.22.-f,73.20.At,79.60.-i}

\maketitle

\section{Introduction}
Single-layer (SL) transition metal dichalcogenides are currently receiving considerable attention mainly due to their electronic similarity to graphene, but with the important distinction of having a seizeable band gap. They constitute a promising materials platform for exploring spin and valley degrees of freedom  \cite{Cao:2012ab,Zeng:2012aa,Xiao:2012ab} and the consequences of strongly bound excitons in two-dimensions \cite{Cheiwchanchamnangij:2012aa,Ramasubramaniam:2012aa,Qiu:2013aa}. Particular focus has been on MoS$_2$ \cite{Mak:2010aa,Splendiani:2010aa,Cao:2012ab,Radisavljevic:2011aa,Zeng:2012aa,Xiao:2012ab}, a material for which SLs can be readily obtained by a micromechanical exfoliation method similar to that used for the initial isolation of graphene. SL MoS$_2$ has  been used in applications such as field effect transistors \cite{Radisavljevic:2011aa} and optoelectronic devices \cite{Lopez-Sanchez:2013aa}. Moreover, MoS$_2$ nanoclusters had already been grown in SLs and used in catalysis even before the advent of graphene \cite{Topsoe:1996aa,Helveg:2000aa,Lauritsen:2003aa,Jaramillo:2007aa}. 

The key difference between the electronic structure of SL MoS$_2$ and its three-dimensional (3D) parent compound is the nature of the band gap. 3D MoS$_2$ in the stable 2H-structure has an indirect band gap with the valence band maximum (VBM) at the $\Gamma$ point of the Brillouin zone (BZ) and the conduction band minimum (CBM) between the $\Gamma$ and $K$ points. SL MoS$_2$, on the other hand, has both the VBM and CBM at the $\bar{K}$ point of the two-dimensional (2D) BZ (corresponding to the $K$\textendash$H$ line in the bulk BZ) \cite{Bollinger:2001aa,Lebegue:2009aa,Mak:2010aa,Splendiani:2010aa,Cheiwchanchamnangij:2012aa}. The band character change between bulk and SL happens mainly because the original VBM and CBM are shifted to lower and higher energy, respectively. In contrast, the states around $\bar{K}$ (or $K$\textendash$H$) are not significantly affected by the structural change and turn into the new VBM and CBM of the SL.

This band gap change can be understood qualitatively by considering the bulk band structure of the material: the bands forming the VBM and CBM of bulk MoS$_2$ show dispersion in the direction perpendicular to the MoS$_2$ sheets of the crystal and are thus subject to quantum confinement for thin crystals (to see this, compare the dispersion of the bands in the $\Gamma$\textendash$K$ and $A$\textendash$H$ directions in e.g. Ref. \cite{Cheiwchanchamnangij:2012aa}). The states forming the VBM and CBM of the SL, on the other hand, show hardly any dispersion in the $K$\textendash$H$ direction and are therefore unaffected by quantum confinement effects. These differences are inherently related to the orbital character of the different states, where higher dispersion in the $K$\textendash$H$ direction indicates a larger contribution from out-of-plane orbitals. In this simple picture, placing a SL of MoS$_2$ onto a substrate can be expected to have a severe impact on the band structure since it strongly modifies the boundary condition for quantum confinement on one side of the SL. Moreover, placing the SL on a metallic substrate  has the potential to alter the band structure and band gap of the SL due to  increased screening from the substrate \cite{Komsa:2012aa}. 

In this article, we explore the band structure and band gap of epitaxial SL MoS$_2$ grown on Au(111). Such growth results in the large-area and high-quality SL materials necessary for applications in devices. Starting from nanoscale clusters of MoS$_2$ on Au(111) \cite{Helveg:2000aa,Lauritsen:2007aa}, we have recently developed a procedure to nearly saturate almost the entire surface of Au(111) by SL MoS$_2$ \cite{Sorensen:2014aa,Gronborg:2015aa} and we have presented initial experimental results of the system's electronic structure determined by angle-resolved photoemission spectroscopy (ARPES) in Ref. \cite{Miwa:2015aa}. Overall, the (occupied) electronic structure of the supported SL closely resembles the theoretical predictions for free-standing SL MoS$_2$.  Details such as the size of the spin-orbit splitting at the $\bar{K}$ point of the BZ are in near quantitative agreement with the predictions \cite{Miwa:2015aa,Zhu:2011ad}. However, some notable differences with respect to free-standing SL MoS$_2$ were observed, suggesting a significant MoS$_2$-substrate interaction. In the experimental part of this paper, we give a complete description of the band structure of MoS$_2$ on Au(111) as determined by ARPES and provide additional data on the band gap size by scanning tunnelling microscopy and spectroscopy (STM and STS). In the theoretical part, we compare these results to first principles calculations. We clarify the nature of the MoS$_2$-substrate interaction, establish a theoretical understanding of the transition metal dichalcogenide-metal interaction and explain the deviations between the ARPES results and the band structure of a free-standing layer.

\section{Methods}

Epitaxial SL MoS$_2$ has been grown on Au(111) by methods described elsewhere \cite{Sorensen:2014aa,Gronborg:2015aa}.  The total MoS$_2$ coverage for most of the ARPES data shown here and in Ref.\cite{Miwa:2015aa} was kept  below one monolayer (ML) at $\approx0.65$~ML in order to avoid the growth of 2~ML islands. The epitaxial SL MoS$_2$ samples are stable in air and could thus be removed from the dedicated growth chamber, transported to the SGM-3 end station of the synchrotron radiation source ASTRID2 \cite{Hoffmann:2004aa} and cleaned via mild annealing to 500~K.  This procedure has been verified to yield atomically clean surfaces by STM \cite{Gronborg:2015aa}. ARPES data were collected at 80~K with an energy and angular resolution better than 20\,meV and 0.2$^{\circ}$, respectively. The measurements presented here were performed at two different photon energies, 49\,eV and 70\,eV.  The dispersion of the highest valence band of MoS$_2$ is most intense in the 49\,eV  data while the 70\,eV data is dominated by the Au bulk bands.  Although not presented in this article, photon energy scans were performed to confirm the lack of $k_z$ dispersion of the SL MoS$_2$ bands to distinguish them from the underlying dispersing Au bulk bands.

STM and STS measurements were carried out using a commercial low-temperature system (CreaTec Fischer \& Co.~GmbH~\cite{Createc}), operating at a base temperature of 6\,K. An etched W tip was used for all measurements, which was flashed \emph{in-situ} and further conditioned by dipping it into the Au substrate prior to the measurements. STS was performed using lock-in detection applied to the sample with a typical frequency range of 741\textendash900\,Hz. 

\begin{figure}
\includegraphics{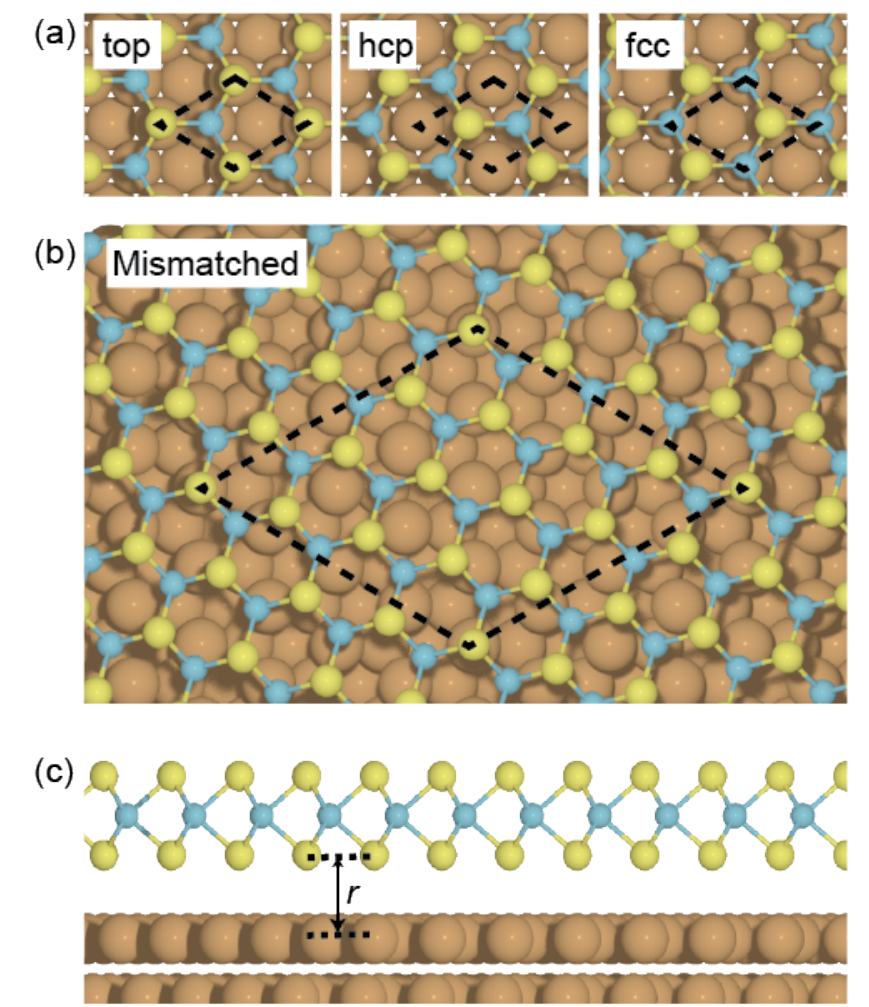}\\
\caption{(Color online) Models used for the DFT calculations. a) Top views of the matched model consisting of a 1$\times$1 unit cell of MoS$_2$ on a 1$\times$1 cell of Au(111). The different positions considered of the bottom S atoms with respect to the underlying Au lattice (on-top, hcp, and fcc) are shown. b) Top view of the mismatched model consisting of a $\sqrt{13}\times\sqrt{13}$ R13.9$^{\circ}$ cell of MoS$_2$ on a 4$\times$4 cell of Au(111). Side view of the MoS$_2$/Au(111) interface indicating the relevant distance $r$ between the lower S layer and the outermost surface layer of Au(111). Turquoise, yellow, and brown spheres indicate the position of Mo, S, and Au atoms, respectively.} 
  \label{fig:1}
\end{figure}

Periodic first principles calculations based on  density functional theory (DFT) were performed for the different SL MoS$_2$ on Au(111) models shown in Fig.\ref{fig:1}: the ``matched" models and the ``mismatched" model. The simple matched models (Fig.\ref{fig:1}(a)) consist of a (1$\times$1) cell of MoS$_2$(0001) on Au(111), with different relative positions of the MoS$_2$ layer with respect to the substrate (the S atoms in the lower layer are located in either the on-top, fcc, or hcp sites of the Au(111) surface). In order to achieve matching of the MoS$_2$ and Au(111) lattices, the lattice parameter of the Au substrate (modelled by 10 atomic layers) was expanded by 9\,$\%$ to  fit the theoretical value for MoS$_2$ (3.18\,\AA). 
Such an expansion results in an artificial destabilization of the Au surface and to shifts of Au surface and bulk bands to higher energies. Although such artifacts indicate that this simple model is inadequate, especially for describing the substrate, it is a useful tool to directly study the MoS$_2$-Au interaction and its effect on the band structure of the SL MoS$_2$. 

We also performed calculations with the more complex and representative mismatched model shown in Fig.\ref{fig:1}(b). The lattice mismatch between SL MoS$_2$ and the underlying Au gives rise to a moir\'e pattern \cite{Sorensen:2014aa,Gronborg:2015aa}. The resulting moir\'e structure consists of a (10$\times$10) supercell of MoS$_2$ on a (9$\times$9) supercell of Au(111). However, since the model required for such structure is computationally unfeasible, our mismatched model consists of a ($\sqrt{13}\times\sqrt{13}$ R13.9$^{\circ}$) supercell of MoS$_2$ supported on a (4$\times$4) supercell of Au(111) (modelled by 4 atomic layers). This model only requires a 0.15\% contraction of the Au lattice and was suggested by Farmanbar \textit{et al.~}\cite{Farmanbar:2015aa} in order to avoid the artificial distortion of the work function of the Au surface. Although it is slightly smaller than the experimentally observed supercell, it shares most of the structural features observed in  experiments, including the different S-Au contact regions - i.e., on-top, hcp, and fcc, within the unit cell.

All the periodic electronic structure calculations presented here were carried out using the VASP code \cite{Kresse:1993v1,Kresse:1996v2,Kresse:1996v3}. The valence electrons were described with plane-wave basis sets with a kinetic energy threshold of 415\,eV, and the interaction between the valence and frozen core-electrons was accounted for by means of the projector-augmented-wave (PAW) method of Bl\"ochl \cite{Blochl:1994}. The PBE approximation to the exchange-correlation functional was used \cite{Perdew:1996}. For the matched models, the reciprocal space was sampled with a (20$\times$20$\times$1) mesh of $k$-points, whereas for the mismatched model, the geometry of the supercell was optimized using a  (4$\times$4$\times$1) mesh, and the charge density was subsequently recalculated with a single point calculation using a denser (10$\times$10$\times$1) mesh of $k$-points. An energy threshold of $10^{-6}$\,eV was used to define self-consistency of the electron density. All atomic positions for the matched models were relaxed until the forces on all atoms were smaller than 0.01\,eV\r{A}$^{-1}$, whereas for the mismatched only the atoms corresponding to the MoS$_2$ layer were relaxed.  The band structure of the matched models can be directly compared to the measured dispersion, but for the larger mismatched supercell, the band structure along the high symmetry directions of the SL MoS$_2$ primitive cell is ``folded" into the smaller reciprocal lattice of the supercell. In order to ``unfold" the band structure with the symmetry of the primitive unit cell of SL MoS$_2$, we have calculated the effective band structure proposed by Popescu and Zunger \cite{Popescu:2012} as implemented in the BandUp code \cite{Medeiros:2014,Medeiros:2015aa}. Spin-orbit coupling has been included for all band structure calculations.

\section{Results and Discussion}

\begin{figure}
\includegraphics{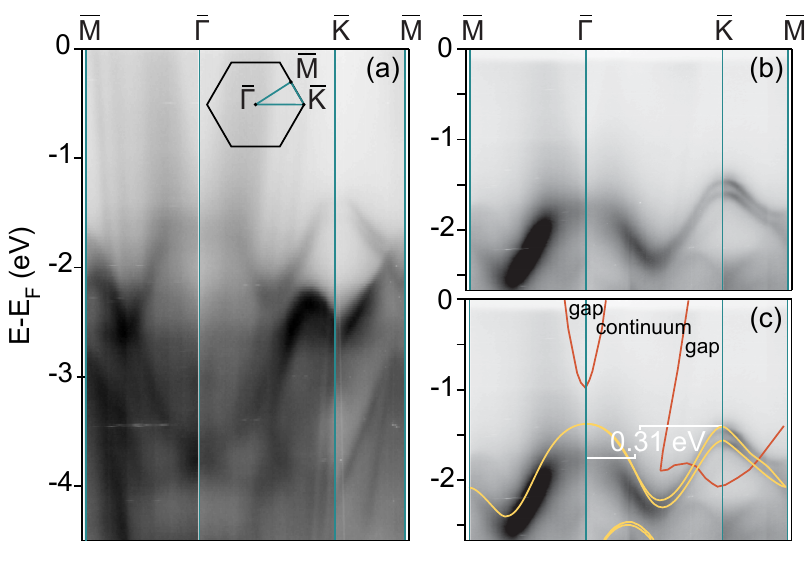}\\
\caption{(Color online) (a) Photoemission intensity for epitaxial SL MoS$_2$ on Au(111) along different high-symmetry directions of the BZ, measured at a photon energy of 70\,eV; inset: hexagonal BZ. (b) Data acquired at a photon energy of 49\,eV in the energy region of the top valence band of MoS$_2$. (c)  The same data as in (b) but with the theoretical dispersion for a free-standing layer (solid yellow lines) and projected band gap edges of the (111) surface of Au (solid orange lines) \cite{Takeuchi:1991aa} superimposed on the experimental data. The measured difference in the top of the valance-band at $\bar{K}$ compared to $\bar{\Gamma}$ is noted to be 0.31\,eV.} 
  \label{fig:2}
\end{figure}

Figure \ref{fig:2} shows the photoemission intensity of epitaxial SL MoS$_2$ on Au(111) along different high symmetry directions in the 2D BZ, measured at two different  photon energies, 70\,eV and 49\,eV (For constant binding energy cuts through the BZ for a photon energy of 49\,eV see Fig. 1(a) of Ref. \cite{Miwa:2015aa}.) The Au substrate contributes with the $sp$ valence band,  the surface state in the L-gap of the bulk BZ (see Supplementary Material of Ref. \cite{Miwa:2015aa}), and the deeper lying $d$-bands visible in the lower part of Fig. \ref{fig:2}(a). 

Projecting the bulk band structure of Au onto the (111) surface gives rise to several projected band gaps \cite{Takeuchi:1991aa}. Wide projected gaps are found near the Fermi energy at the $\bar{\Gamma}$, $\bar{K}$ and $\bar{M}$ points. These gaps appear as a reduction of background intensity and are outlined by the solid orange lines in Fig. \ref{fig:2}(c).

The upper valence band of SL MoS$_2$ is clearly identified in the measurements acquired with 49\,eV photons in the energy range between (-1.39$\pm$0.03)\,eV (top of the valence band at $\bar{K}$) and (-2.49$\pm$0.03)\,eV (bottom of the valence band between $\bar{\Gamma}$ and $\bar{M}$). 
For comparison, the calculated upper valence band for free-standing SL MoS$_2$ is superimposed on the photoemission data as yellow lines in Fig. \ref{fig:2}(c). The overall agreement between experiment and this calculation is excellent. In particular, the dispersion close to $\bar{K}$ and the size of the spin-splitting there is reproduced almost quantitatively. But there are also a few points of disagreement. The first is the relative energy of the local valence band maxima at $\bar{\Gamma}$ and $\bar{K}$. In the calculation for the free-standing layer, these states are almost degenerate whereas the state at $\bar{\Gamma}$ is measured to be at significantly higher binding energy ((-1.71$\pm$0.03)\,eV) than the state at $\bar{K}$ for SL MoS$_2$ on Au(111). Following the arguments given in the introduction, it is reasonable to assume that this (0.31\,eV)  change is due to the boundary condition given by the MoS$_2$-Au(111) interface and we shall see below that this is indeed the case. Note that such a distortion of the uppermost valence band does not take place for exfoliated SL MoS$_2$ deposited on SiO$_2$ substrates \cite{Jin:2013aa}, presumably due to the weaker bonding of the sulphur atoms to the substrate. 

A second point of disagreement between the measured dispersion and the calculation for the free-standing layer is the dispersion of the bands around $\bar{M}$. In the calculation, the spin-splitting of the bands reduces to zero at $\bar{M}$. This  is expected at $\bar{M}$ (and at $\bar{\Gamma}$) due to the combination of time-reversal and crystal symmetry. $\bar{M}$ is a so-called time-reversal invariant momentum in the 2D BZ \cite{Teo:2008aa}, meaning that two $\bar{M}$ points can be connected both by an inversion of the $\mathbf{k}$-direction and by a reciprocal lattice vector. The combination of these symmetries enforces a spin-degeneracy in the 2D electronic states. This degeneracy is observed at $\bar{\Gamma}$ but the situation is somewhat unclear at $\bar{M}$. We shall return to these two points of disagreement below when we discuss the calculations for SL MoS$_2$ adsorbed on Au(111). 

The ARPES data only gives very limited information on the character and size of the SL MoS$_2$ band gap since only states below the Fermi energy are accessible. As a consequence, we can only conclude that the band gap must be larger than $\approx$1.3\,eV, i.e. corresponding to the energy of the VBM. We have previously attempted to determine this band gap by alkali atom doping \cite{Miwa:2015aa} and by time- and angle-resolved photoemission spectroscopy, i.e. by pumping electrons into the conduction band and subsequent emission by a second photon \cite{Cabo:2015}. Adsorbing potassium atoms onto the surface does indeed lead to a sufficiently strong charge donation to push the CBM below the Fermi level, resulting in the observation of a direct band gap at $\bar{K}$ with a size of (1.39$\pm$0.05)\,eV. However, it is questionable if this measured gap corresponds to the gap of the pristine system as potassium adsorption also leads to a distortion of the occupied bands in which states with out-of-plane orbital character (e.g. the upper valence band at $\bar{\Gamma}$) are shifted more than states with in-plane orbital character (e.g. the VBM at $\bar{K}$). Indeed, it has been found that sodium intercalation in bulk MoS$_2$ does not only lead to electron donation but to a change of the band gap as such \cite{Komesu:2014ab} and can even trigger a transition from the semiconducting 2H-phase to the metallic 1T-phase \cite{Wang:2014acs}. The band gap determined by pump/probe ARPES, which does not require doping with alkali metals, is (1.95$\pm$0.05)\,eV \cite{Cabo:2015}.

It is important to note that this electronic band gap is a fundamentally different quantity from the optical gap, which is generally measured by absorption and photoluminescence experiments  \cite{Mak:2010aa,Splendiani:2010aa}. In simple terms, the optical band gap corresponds to the energy required to create an exciton while the electronic band gap also requires the breaking of the exciton and is thus higher due to the exciton binding energy. While the electronic band gap for free-standing SL MoS$_2$ is not known, an optical band gap of $\approx 1.9$\,eV has been determined for free-standing SL MoS$_2$~\cite{Mak:2010aa} as well as for SL MoS$_2$ on graphite~\cite{Zhang:2014ae}, for which weaker interactions than for Au(111) are expected. This optical gap measurement is in excellent agreement with a quasi-particle calculation of the optical gap of a free-standing layer \cite{Qiu:2013aa}, which in addition predicts an electronic gap of 2.8\,eV.

\begin{figure}
\includegraphics{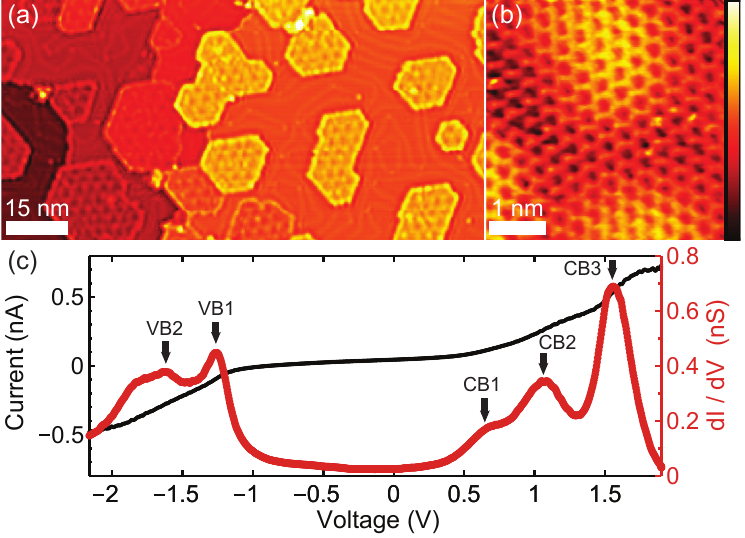}\\
\caption{(Color online) (a) Constant-current STM image of 0.5\,ML coverage of SL MoS$_{2}$ on Au(111).  Image parameters: V$_{S}$ = $-$1.5\,V, I$_{t}$ = 0.2\,nA.  (b) Atomically resolved image of a SL MoS$_{2}$  island revealing the atomic lattice and  moir\'e pattern.  Image parameters: V$_{S}$ = $-$0.12\,V, I$_{t}$ = 0.1\,nA. (c) Representative STS (current and differential conductance ($dI/dV$)) near the center of a SL MoS$_{2}$ island. The tip was stabilized at V$_{S}$ = 1.9\,V, I$_{t}$ = 0.4\,nA. A modulation amplitude of 42\,mV was used.}
  \label{fig:3}
\end{figure}

Another reliable approach for determining the electronic band gap of a material is STS. We therefore performed STS measurements on numerous MoS$_2$ islands to simultaneously determine the electronic structure in occupied and unoccupied regimes as well as its spatial dependence. Figure \ref{fig:3} shows STM and STS data from the sample. Acquiring position-dependent STS spectra on the islands (which correspond to the brighter patches with the distinct  moir\'e pattern in Fig. \ref{fig:3}(a)), we observe small spatial variations within an energy range of 200 meV. However, we found no statistical correlation between variations in intensity or peak location and the spatial location within the unit cell of the moir\'e pattern. Therefore, we only considered spatially averaged spectra taken near the center of MoS$_2$ islands and away from domain boundaries (edges) for determining the energy gap. STS acquired in a $\pm$2\,eV energy window (Fig.~\ref{fig:3}(c)) shows a regime of very low conductance around the Fermi energy, i.e. between $-$1\,eV and 0.5\,eV, with onsets of the conductance at higher absolute energies. The STS spectra greatly differ from those acquired on the clean Au(111) surface areas which only show the typical surface state onset below the Fermi energy. STS spectra of the  SL MoS$_2$ islands show clear conductance features resulting from the emergence of SL MoS$_2$ bands at different energies -- these are indicated by arrows in Fig.~\ref{fig:3}(c). The sharp rise in conductance is due to the onsets of the SL MoS$_2$ valence and conduction bands outside of the MoS$_2$ band gap. In similar STS experiments involving SL MoSe$_2$ on graphene, features in the $dI/dV$ versus $V$ plot are harder to discern but it is possible to determine the exact position of the band edges by plotting $\log(dI/dV)$ versus $V$ instead \cite{Ugeda:2014aa,Bradley:2015}. In the present case, the metallic character of the support and its strong hybridization with SL MoS$_2$ lead to a net (albeit low) conduction in the band gap and to smoother conductance variations, which hinders the possibility of assigning small increases to the band edges. We therefore approximate onsets to the peaks in the STS spectra. 

In the valence band regime, we can directly compare the resonances at VB1\,=\,$(-1.24\pm0.06)\,$eV and VB2\,=\,$(-1.60\pm0.06)\,$eV to the top of the valence band at $\bar{K}$ and $\bar{\Gamma}$ in the ARPES data shown in Fig.~\ref{fig:2}(c).  The energy difference between VB2 and VB1 is 0.36\,eV and in agreement with the $\approx$0.31\,eV determined from ARPES data for the same sample. In the conduction band, we see three onsets (CB1-3), specifically a  step centered at $(0.50\pm0.27)$\,eV (CB1) and two resonances at $(0.99\pm0.06)$\,eV (CB2) and $(1.46\pm0.06)$\,eV (CB3). Correlating CB1 and VB1 to the $\bar{K}$ band onsets yields a band gap value of (1.74$\pm$0.27)\,eV, where the larger error stems from the uncertainty in defining the position of the CB1 step.

While the gap agrees reasonably well with the pump/probe ARPES results of Ref. \cite{Cabo:2015}, STS measurements on similar samples, but carried out at room temperature, have been reported to yield a smaller band gap of 1.3\,eV \cite{Sorensen:2014aa}. This difference can most likely be ascribed to the different experimental temperature (6\,K here vs room temperature in Ref. \onlinecite{Sorensen:2014aa}) that affects the size of the gap as such (the spectra differ mainly in the position of VB1) but also has a significant impact on stability and energy resolution. The smaller band gap (1.74\,eV) measured here for MoS$_2$ on Au in comparison to the 2.15\,eV measured for MoS$_2$ on graphite using STS \cite{Zhang:2014ae} supports the interpretation of a  stronger band structure renormalization induced by the Au(111) support \cite{Cabo:2015}.

\begin{figure}
\includegraphics{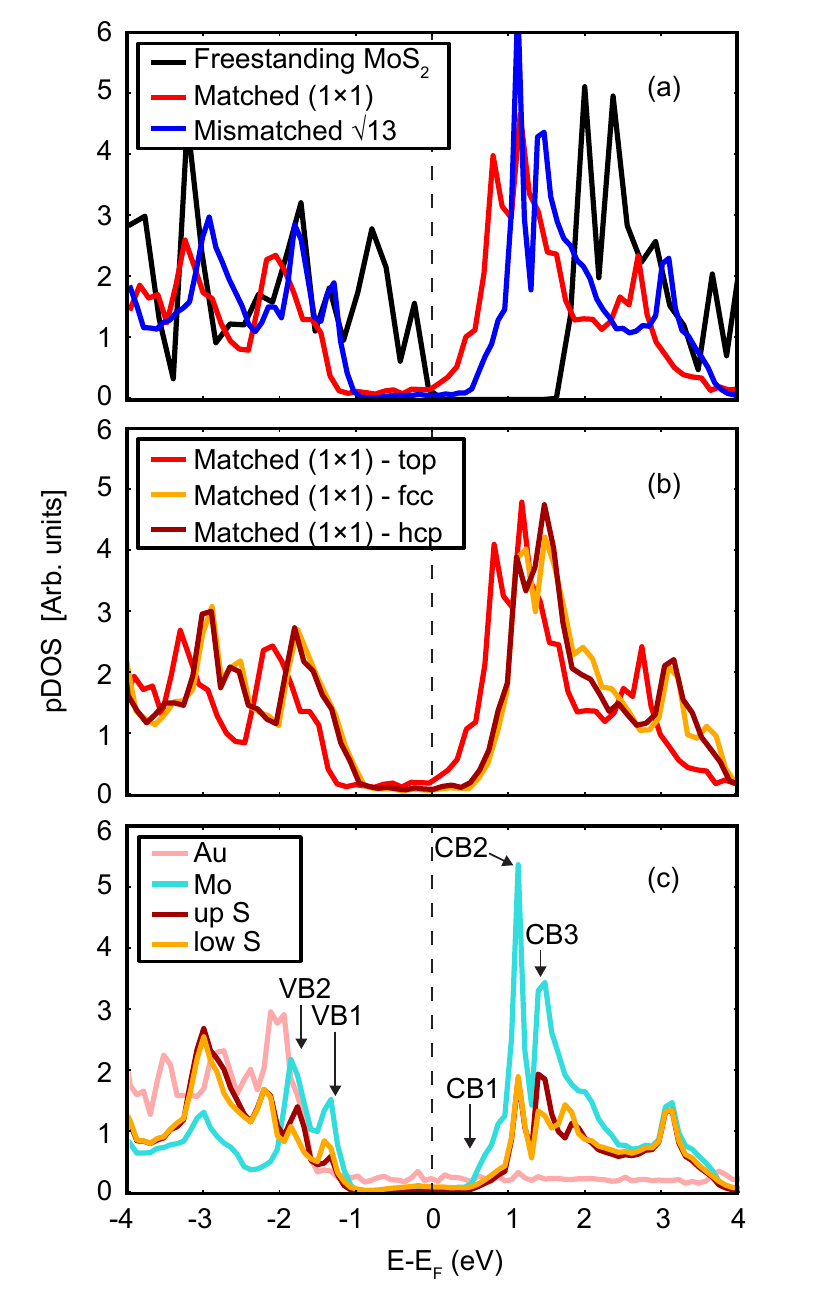}\\
\caption{(Color online) (a) Calculated projected density of states (pDOS) of the MoS$_{2}$ layer in the matched (on-top) and mismatched MoS$_2$/Au(111) models and for the free-standing monolayer. (b) pDOS on MoS$_2$ for the on-top, hcp, and fcc positions for the matched model. (c) DOS projected on the different atom types of the mismatched model, i.e. Au, Mo, and S atoms from the upper and lower layers. The pDOS of the S atoms has been scaled $\times$3 for clarity. Peaks in (c) have been labeled according to the assignments made in Fig.~\ref{fig:3}. The onsets of VB1 and CB1 are -1.05\,eV and 0.54\,eV, respectively. The peaks of VB2, CB2, and CB3 are -1.83\,eV, 1.15\,eV, and 1.43\,eV, respectively.}
  \label{fig:4}
\end{figure}

The experimental results on the geometry and electronic structure are now compared to first principles calculations based on DFT. We have optimized the geometry for the different models of a SL MoS$_2$ on Au(111)  in Fig.~\ref{fig:1}  (three matched models for adsorption in different sites and a mismatched model). A structural optimization of the matched models results in the on-top position model being the most stable one by 199\,meV and 215\,meV per unit cell, compared to the hcp and fcc models, respectively. This higher stability is accompanied by a smaller optimized distance between the lower S layer and the outermost Au layer (distance $r$ in Fig~\ref{fig:1}(c)) for the on-top structure (2.51\,\r{A}), as for the fcc and hcp structures (3.20 and 3.03\,\r{A}, respectively).  Interestingly, such height differences are in close to quantitative agreement with those measured in experimental STM images of SL MoS$_2$ on Au(111)\cite{Sorensen:2014aa}. However, the intuitive assignment made in ref.~\cite{Sorensen:2014aa} of the different areas within the moir\'e pattern, where topographically higher regions are assigned to on-top bonding modes, may be called into question in view of the present results, where the on-top configuration leads to the smallest $r$ of the matched models.
For the mismatched model the optimized $r$ is 3.29\,\r{A}, which is larger than for the matched models, probably as a result of S atoms being, on average, in less favorable positions.

Fig.~\ref{fig:4}(a) show the density of states projected onto the MoS$_2$ layer  (pDOS) for a free-standing layer, a matched layer in the optimum position (on-top) and the mismatched layer, whereas Fig.~\ref{fig:4}(b) compares the pDOS for the different matched models.  Despite of the structural  differences, the overall shape of the pDOS for the different systems is very similar and only the relative position with respect to the Fermi level differs. The different strength of the SL MoS$_2$ interaction with the Au surface leads to binding energies (calculated as the difference in energy between the MoS$_2$-Au system and the separated MoS$_2$ and Au) of $-$0.25, $-$0.03, and $-$0.04 eV for the on-top, fcc, and hcp configurations, respectively. 

In a generic semiconductor, the position of the Fermi level, or rather the chemical potential, is dictated by  band structure, doping and temperature. In the present situation, the interface to the underlying Au plays an important role for the position of the MoS$_2$ bands. In all cases studied here, the interaction between MoS$_2$ and Au(111) gives rise to a $n$-type contact with the Fermi level pinned near the conduction band of MoS$_2$, which is in agreement with similar studies \cite{Popov:2012aa,Kang:2014a,Gong:2014aaa,Farmanbar:2015aa}. However, the resulting energy of the VBM with respect to the Fermi level (E$_F$) differs for the cases considered in this work and for the different models used in the literature. Such shifts of the band alignment can be caused by the different work functions of the underlying metal  substrates \cite{Gong:2014aaa,Farmanbar:2015aa}. Work function differences for the same kind of metal surfaces can be induced by strain and we do in fact find marked differences between the position of the VBM calculated for matched and mismatched models: For the matched model used here and in Ref.~\cite{Popov:2012aa}, which involves a 9~\% Au(111) expansion, the CBM of MoS$_2$ is almost touching the Fermi level. On the other hand, for our mismatched model and the ones used in Refs.~\cite{Kang:2014a,Gong:2014aaa}, which involve a compressive strain of the Au(111) lattice, the CBM is found at higher energies with respect to the E$_F$ (0.6$\textendash$0.8\,eV). Since the mismatched model involves only very little strain of the Au 
 lattice (0.15\%), it should give a more realistic description. In fact, this small strain was one of the reasons why the mismatched model  was originally suggested in Ref. \onlinecite{Farmanbar:2015aa}.

The pDOS results for the different matched models in Fig.~\ref{fig:4}(b) indicate that the Fermi level pinning not only depends on the work function of the Au(111) surface, but also, and even more strongly, on the interaction with the MoS$_2$ layer. This interaction was reported to lead  to the formation of interface dipoles due to charge redistribution and to the formation of gap states with Mo $d$-orbital character \cite{Gong:2014aaa}. Here, we also observe such charge redistribution, which is strongly localized at the interface and extends to the S-Mo-S layer but not to the vacuum region (not shown).

For the on-top matched model the stronger interaction with the surface shifts the MoS$_2$ states towards lower energies than for the other sites. This is due to the more favorable S-Au bonds, which involve stronger charge transfer and redistribution at the interface. For the matched models with the hcp and fcc positions, the interaction and the resulting pDOS shifts are weaker but they are similar to each other, consistent with a similar S-Au bonding geometry. The S-Au bonding for the mismatched model, on the other hand, is on average a mixture of top, hcp, and fcc bonding sites. As a result, the position of the VBM with respect to the E$_F$ for this system is very similar to the hcp and fcc cases despite the differences in work function of the underlying Au substrate models. The small strain of the underlying Au(111) and the different S-Au bonding regions make the mismatched model more representative of the experimental situation. Its DOS is thus more suitable for comparison to the onset values with respect to the Fermi level determined by STS and ARPES. We have therefore labeled the bands in the atomically decomposed pDOS spectra of the mismatched model according to the labels used for the STS spectra (Fig.~\ref{fig:4}(c)). 

The VB1 and CB1 have contributions mainly from the Mo atoms, whereas S atoms also contribute to the VB2, CB2, and CB3. In addition, VB2 overlaps with the $d$-band of Au, which, as we show in the calculated band structures below, leads to a strong hybridization between S and Au states. The onset and peak values of the different bands agree quite well with the STS values extracted from Fig.~\ref{fig:3}(a); the calculated (experimental) onsets of VB1 and CB1 are -1.05\,eV and 0.56\,eV (-1.24\,eV and 0.50\,eV), respectively, whereas  the peaks of VB2, CB2, and CB3 are at -1.83\,eV, 1.15\,eV, and 1.43\,eV (-1.60\,eV, 0.99\,eV, and 1.63\,eV), respectively. The over- and underestimation of the position of VB1 and CB2, respectively, result in a band gap (1.61\,eV) that is underestimated by 0.13\,eV compared with the value determined by STS ((1.74$\pm$0.27)\,eV).

\begin{figure*}
\includegraphics{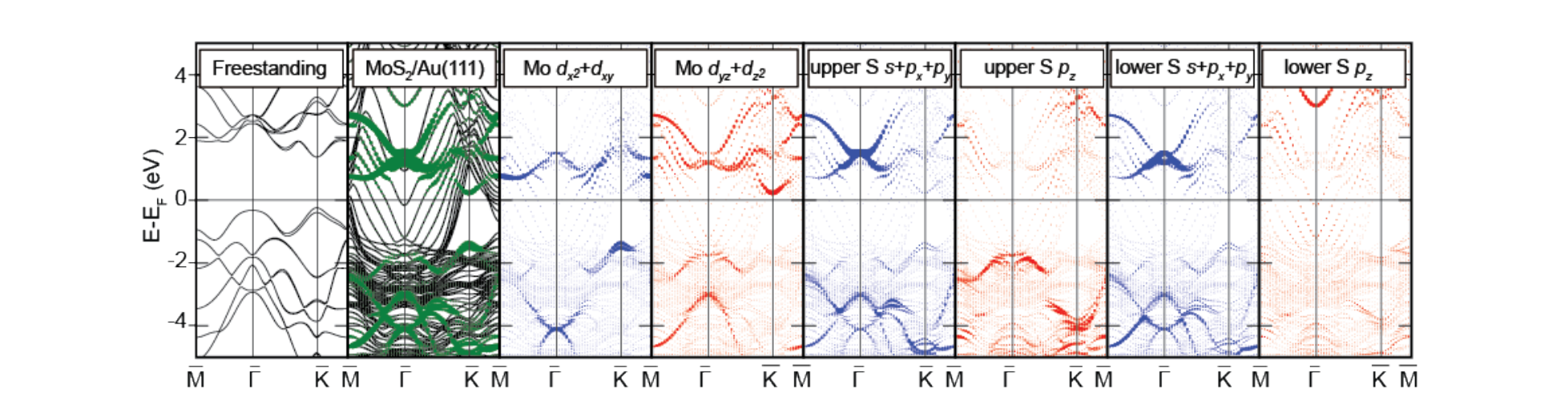}\\
\caption{(Color online) Calculated band structures for free-standing SL MoS$_2$ and SL MoS$_2$ on Au(111) (matched model). The size of the green circles in the second panel indicate the weight for each state from MoS$_2$ orbitals. For the rest of panels, the blue (red) circles indicate the weight from in-plane (out-of-plane) orbitals for the Mo atom and for the sulphur atoms on the MoS$_2$-vacuum (upper S) and MoS$_2$-Au(111) interfaces (lower S). The weight for the S orbitals have been multiplied by two in order to visualize their contributions more clearly.}
  \label{fig:5}
\end{figure*}

The position and character of the bands can be seen more clearly by analyzing the calculated band structures for free-standing SL MoS$_2$ and the SL adsorbed on Au(111) shown in Fig.~\ref{fig:5} (first and second panels). The band structure for the SL on Au(111) corresponds to the matched model in the most stable on-top binding geometry (Fig.~\ref{fig:1}). The calculated band structure for the free-standing layer is in excellent agreement with previous results using the same exchange-correlation functional \cite{Zhu:2011ad}, and the resulting band gap (1.58\,eV) is also very similar.  This value significantly underestimates the electronic gap of ~2.8\,eV determined by more sophisticated quasi-particle (GW) calculations \cite{Qiu:2013aa}.  Interestingly, calculations with the local density approximation (LDA)\cite{Kang:2014a,Gong:2014aaa} or with another exchange-correlation functional (PW91) within the generalized gradient approximation (GGA) lead to slightly ($\approx 0.2$\,eV) larger values \cite{Li:2007} than the 1.58~eV calculated here. Band gap underestimation is a well-known shortcoming of the LDA and GGA approaches, which can be partially remedied by including a fraction of Hartree-Fock exchange, leading to the so-called hybrid functional formalism. Indeed, calculations with such functionals yield improved band gaps of ~2.05\,eV \cite{Ramasubramaniam:2012aa,Padilha:2014a}. Nevertheless, studying the MoS$_2$/Au(111) with standard DFT methods provides valuable insight into the effect of the substrate on the band structure despite the gap underestimation. 

For the on-top matched model, the adsorption of the MoS$_2$ layer on Au(111) leads to pronounced changes in its band structure (Fig.~\ref{fig:5}, second panel) but, surprisingly, the fundamental gap near $\bar{K}$ (1.58\,eV) is only 0.02\,eV lower than for the free-standing layer. Moreover, the bands forming the VBM and CBM of the layer remain well-defined, something that is ascribed to their greater contribution from Mo orbitals (which do not directly bond to the Au substrate), to the in-plane nature of such states for the VBM and, most importantly, to their position in a projected gap of the Au(111) band structure. The binding energies of the occupied bands for the matched model are generally in good agreement with the experimental values from ARPES, and the position in the projected band gap is consistent with the VBM states being very sharp in ARPES. However, the exact energy of the bands for the matched model is affected by the artificial strain induced in the Au(111) slab, as described above, and should thus not be directly compared to the ARPES values. The unchanged size of the band gap with respect to the free-standing SL MoS$_2$, on the other hand, is a common feature of all models, and the same was found for other mismatched models with compressed Au(111) surfaces when studied with similar levels of theory~\cite{Kang:2014a,Gong:2014aaa}. 

In contrast to the situation around the fundamental gap, the upper valence band near $\bar{\Gamma}$ is strongly affected by the adsorption on Au(111). The band is still well-defined as it enters the bulk continuum and its spin-splitting is lost. However, very close to $\bar{\Gamma}$ it strongly merges with the bulk bands. The band maximum at $\bar{\Gamma}$ is very diffuse but significantly lower (by $\approx$0.4\,eV) than at the $\bar{K}$ point. These observations are in excellent agreement with the experiment where a similar energy shift and a strong broadening, indicative of MoS$_2$-substrate interaction, is observed. In fact, this band structure change is consistent with the simple expectation that the adsorbate-substrate interaction should mainly affect the boundary condition for the quantization of the out-of-plane states in the SL. Note that the MoS$_2$ character is retained for higher energy states near $\bar{\Gamma}$ and the resulting dispersion of these states is very similar to the ARPES results in Fig.~\ref{fig:2}(a). 

The band structure can be further explored by decomposing the states of the SL MoS$_2$/Au(111) model into the contribution from the Mo atoms and from the lower (in contact with Au) and upper (in contact with the vacuum) layer of S atoms. Fig.~\ref{fig:5} shows such weighted band structures, separated into out-of-plane orbitals and in-plane orbitals. The VBM of the MoS$_2$ layer is mostly formed from in-plane Mo 4$d$ states at the $\bar{K}$ point (see also Refs. \cite{Zhu:2011ad,Cappelluti:2013aa}) which, together with their position in the projected band gap, explains why it remains relatively unaffected by the interaction with the Au(111) surface. The CBM at the $\bar{K}$ point, on the other hand, is formed by out-of-plane Mo 4$d$ states, which are expected to be distorted by out-of plane interactions. Considering this and the low gap measured by STS, a gap variation upon interaction with the Au substrate is expected. Nevertheless, the size of the fundamental gap at $\bar{K}$ calculated here is unaffected by the interaction with the metallic substrate. This points to the inefficiency of the semi-local LDA or GGA approaches in capturing subtle effects of out-of-plane interactions in band gaps involving correlated states. These methods suffer from an inherent self-interaction error and only partially account for electronic
correlation, neglecting long-range exchange. In contrast, GW calculations account for long-range Coulomb interaction and are able to predict subtle screening effects. For SL MoSe$_2$ (which has a very similar electronic structure to MoS$_2$) a small direct gap reduction is calculated upon interaction with a support \cite{Ugeda:2014aa}, and lower $\bar{K}$-$\bar{K}$ gaps are predicted when increasing the number of layers or decreasing the interlayer distance in MoS$_2 $\cite{Komsa:2012aa}. 

In contrast to the situation at $\bar{K}$, the VBM at $\bar{\Gamma}$ contains significant contributions from the sulphur $3p_z$ states (as well as some contribution for out-of-plane Mo 4$d_{z^{2}}$ and 4$d_{yz}$ orbitals) and here we observe a strong asymmetry for the top and bottom sulphur atoms. While the top-atoms still participate in the formation of the upper valence band states at $\bar{\Gamma}$, the contribution of the $3p_z$ orbitals of the bottom sulphur atoms is completely hybridized and merged with the Au(111) states. As one might have expected, this is an indication of a strong Au-S bonding in the system. This strong bond will presumably limit the possibilities to modify the electronic properties of epitaxial MoS$_2$ on Au(111) compared to the case of graphene on transition metal surfaces, where a large number of atomic species can be intercalated between the graphene and the surface in order to change the graphene's electronic properties \cite{Varykhalov:2008aa,Lizzit:2012aa,Larciprete:2012aa,Ulstrup:2014ad,Petrovic:2013aa}. 
 
The strong interaction with the bulk state continuum can also explain the failure to observe the expected spin-degenerate upper valence band at $\bar{M}$. As can be seen in Fig.~\ref{fig:5}(second panel), the upper valence band mixes so strongly with the Au states that it does not exist as a well-defined state at $\bar{M}$. This is in contrast to exfoliated SL MoS$_2$ deposited on SiO$_2$ where the upper valence band can be observed in the entire 2D BZ, albeit without a well-resolved spin-orbit splitting \cite{Jin:2013aa}. 

\begin{figure}
\includegraphics{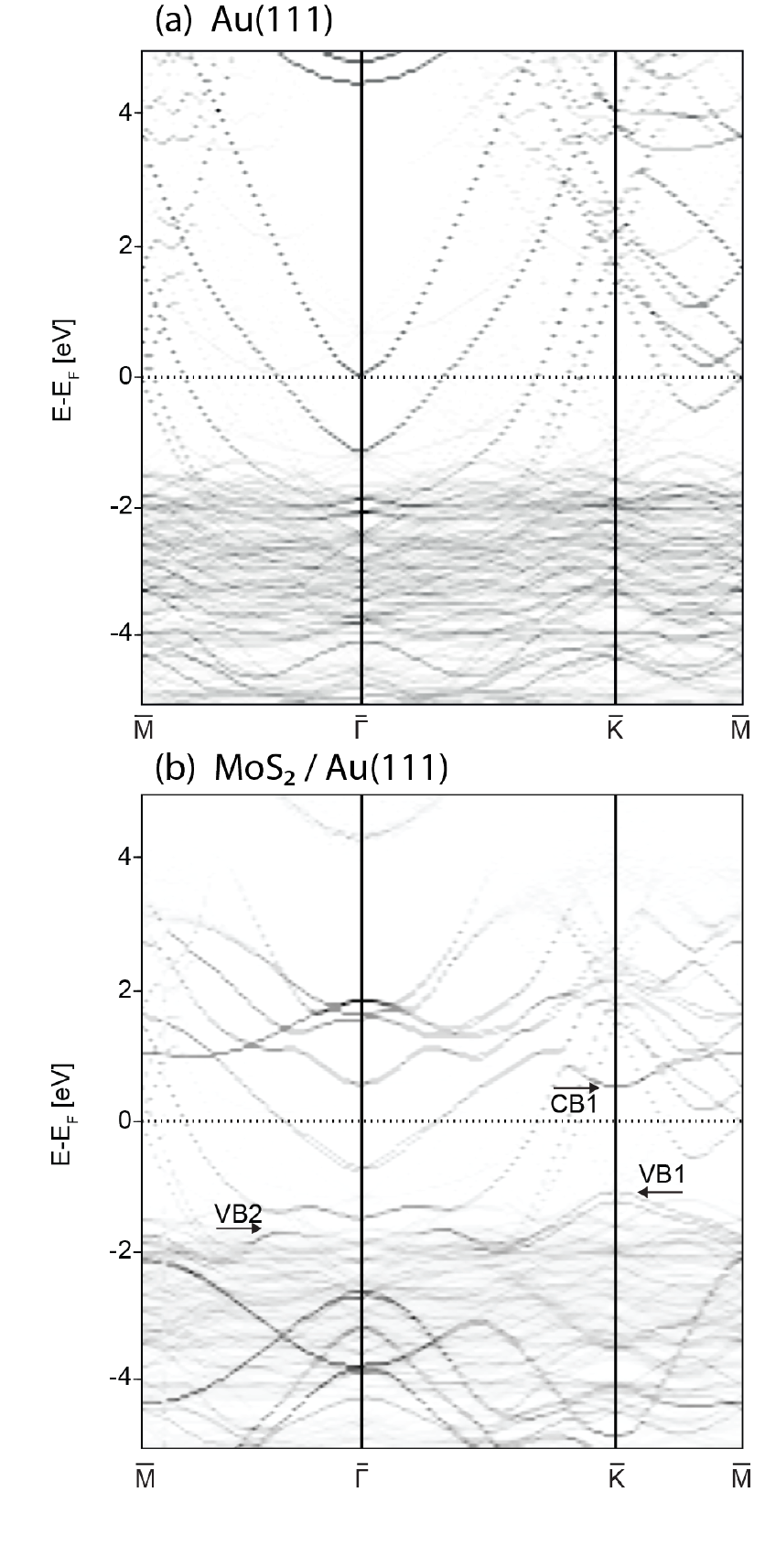}\\
\caption{ Unfolded bandstructures for (a) a bare 4$\times$4 Au(111) surface and (b) the mismatched model consisting of a $\sqrt{13}\times\sqrt{13}$ R13.9$^{\circ}$ cell of MoS$_2$ on a 4$\times$4 cell of Au(111). The position of relevant MoS$_2$ bands (VB1, VB2, and CB1) has been indicated according to the labels in Figs.~\ref{fig:3}(a) and \ref{fig:4}(c).}
  \label{fig:6}
\end{figure}

Let us now compare these results to the band structure of the mismatched model. As mentioned in the introduction, the band structure of such a supercell is folded into its smaller BZ. In order to recover the primitive cell picture of the band structure of Au-supported MoS$_2$, we have calculated the so-called effective band structure (EBS) for this system. To help differentiate between MoS$_2$ and Au bands, Fig.~\ref{fig:6}(a) shows the band structure for the Au(111) slab (without the MoS$_2$ layer) unfolded on the primitive cell of MoS$_2$(0001).The intensity in the EBS plots for an energy interval $dE$ and a $k$-vector of the primitive cell depends on the number of states of the supercell that have the same character as the primitive cell k-vector in that energy interval, i.e. that are related by the unfolding operator \cite{Medeiros:2014,Medeiros:2015aa}. 

Unfolding the band structure of the Au(111) supercell on the primitive cell of MoS$_2$(0001) results in an EBS with several bands, which shows that despite the mismatch, there are states of the Au(111) slab that have a similar character as the primitive cell of MoS$_2$(0001). In particular, the $d$-band continuum of Au appears diffuse below $\approx$-2.0\,eV and several bands cross the Fermi level, some of which have minima at $\bar{\Gamma}$. This suggests that one of these bands - probably the one with the minima closest to the Fermi level - corresponds to the surface state of Au(111). However, the  energy of the surface state of Au(111) is not well-reproduced and this is due to the model used. For this mismatched model, the Au(111) surface is represented by just four atomic layers whose structure is fixed to their truncated bulk positions. Using a much thicker Au(111) slab model and relaxing its structure would correctly reproduce the surface state, but such calculations would be too computationally demanding.  In addition, it should be noted again that the unfolding is done on the unit cell of MoS$_2$(0001) and this means that the unfolded band structure does not exactly correspond to that of the Au(111) primitive cell.

The EBS for the MoS$_2$/Au(111) mismatched model in Fig.~\ref{fig:6}(b) shares most of the VBM features of the band structure of the on-top matched model. The position of relevant MoS$_2$ bands has been indicated in Fig~\ref{fig:6}(b) by using the same labels (VB1, VB2, and CB1) as in in Figs.~\ref{fig:3}(a) and \ref{fig:4}(c), to help differentiating between Au and MoS$_2$ states. The characteristic spin-split (by 151\,meV) bands at $\bar{K}$ (VB1) are clearly visible, whereas bands with lower energy (VB2) merge with the Au $d$-band, becoming more diffuse closer to $\bar{\Gamma}$ and practically disappear. The same states found crossing the Fermi level in the EBS of the bare Au(111) can also be recognized in the presence of the MoS$_2$ layer, but shifted towards higher energies. This is not surprising given the n-doping of the MoS$_2$ layer, which leads to a concomitant p-doping of Au. Interestingly, the interaction of VB2 with one of the Au bands not only shifts the latter towards higher energies but also leads to avoided crossings between these bands near $\bar{\Gamma}$. In addition, the mixing of VB2 with Au states at $\bar{\Gamma}$ destroys the local maximum of VB2 at $\bar{\Gamma}$. This indicates that the interaction with the Au substrate leads to VB2 having two local maxima near $\bar{\Gamma}$ instead of just the maximum right at $\bar{\Gamma}$ characteristic of free-standing SL MoS$_2$. This is consistent with the ARPES spectra, where the VBM is hardly discernible at $\bar{\Gamma}$. Furthermore, the band gap at $\bar{K}$ from the EBS is 1.61\,eV, which is very similar to those calculated for the free-standing monolayer (1.60\,eV) or for the on-top matched MoS$_2$/Au(111) model (1.58\,eV). It is therefore clear that the choice of model does not affect the fundamental band gap at $\bar{K}$. 

\section{Conclusions}

In summary, ARPES and STS and first principles calculations have been used to study the electronic structure of epitaxial single-layer MoS$_2$ on Au(111). We find that the interaction with a metallic support strongly distorts the VBM at $\bar{\Gamma}$ and reduces the fundamental band gap at $\bar{K}$ with respect to a free-standing layer. The resulting band gap is measured by STS to be ($1.74\pm0.27$)\,eV, which is in relatively good agreement with the ($1.95\pm0.05$)\,eV determined by time-resolved ARPES \cite{Cabo:2015}. Our DFT calculations yield a smaller band gap of 1.61~eV and reveal that the Mo 4$d$ states forming the VBM at $\bar{K}$ of the single layer do not hybridize with the Au(111) substrate. This is due to the position of the MoS$_2$ VBM which does not overlap with the Au bands, and the central position of the Mo atoms in the MoS$_2$ layer. In turn, the formation of Au-S bonds leads to a strong hybridization of the S 3$p_z$ orbitals with the $d$ band of Au, which causes a distortion of the MoS$_2$ VBM at $\bar{\Gamma}$. 

\section{Acknowledgements}
We gratefully acknowledge financial support from the VILLUM foundation, the Danish Council for Independent Research, Natural Sciences under the Sapere Aude program (Grant Nos. DFF-4090-00125, DFF-4002-00029, and 0602-02566B), the Lundbeck Foundation,The Innovation Fund Denmark  (CAT-C) and Haldor Tops\o e A/S. AB acknowledges support from the European Research Council under the European Union's Seventh Framework Programme (FP/2007-2013) / Marie Curie Actions / Grant no. 626764 (Nano-DeSign). NH, DW, and AAK acknowledge financial support from the Alexander von Humboldt foundation, the Emmy Noether
Program (KH324/1-1) via the Deutsche Forschungsgemeinschaft, and from FOM which is part of NWO, and the NWO Vidi program. The authors are grateful to Dr. Paulo V. C. Medeiros for the open-source distribution of the BandUp code and for the useful technical support.  


\end{document}